\documentclass[12pt]{iopart}
\usepackage{graphicx}

\begin{document}

\title[Controlling Excitations Inversion of a CPB Interacting with a NR]{%
Controlling Excitations Inversion of a Cooper Pair Box Interacting with a
Nanomechanical Resonator}
\author{C. Valverde$^{1,\dagger,\ddagger}$, A.T. Avelar$^{\dagger}$ and B.
Baseia$^{\dagger}$}

\address{$^{1}$Universidade Estadual de Goi\'as, Rod. BR 153, 3105, 75132-903 An\'apolis, GO, Brazil.}
\address{$^{\dagger}$Instituto de F\'{\i}sica, Universidade Federal de Goi\'as, 74001-970 Goi\^ania, GO,
Brazil.}
\address{$^{\ddagger}$Universidade Paulista, Rod. BR 153, km 7, 74845-090 Goi\^ania, GO, Brazil.}
\ead{valverde@ueg.br}

\begin{abstract}
We investigate the action of time dependent detunings upon the excitation
inversion of a Cooper pair box  interacting with a nanomechanical resonator.
The method employs the Jaynes-Cummings model with damping, assuming
different decay rates of the Cooper pair box and various fixed and
t-dependent detunings. It is shown that while the presence of damping plus
constant detunings destroy the collapse/revival effects, convenient choices
of time dependent detunings allow one to reconstruct such events in a
perfect way. It is also shown that the mean excitation of the nanomechanical
resonator is more robust against damping of the Cooper pair box for
convenient values of t-dependent detunings.
\end{abstract}

\pacs{03.65 -w, 03.65 Yz, 85.85. +j}



\section{Introduction}

A popular and exactly soluble model in quantum optics is the Jaynes-Cumming
model (JCM). It describes the interaction of a two-level atom with a
single-mode of the electromagnetic field \cite{1,2,2.1,2.2,2.3,2.4,2.5}.
Over the last two decades various extensions of the ordinary JCM have been
used in various directions, e.g., as adapted to: (i) the study of
interaction of a three-level atom with a two-mode squeezed vacuum \cite{10};
(ii) the study of atom-field interaction in the presence of a cavity damping
\cite{11}; (iii) the same as in (i), including an additional (nonlinear)
Kerr medium \cite{12}; (iv) the two-level atoms inside a cavity acted upon
by an external field control \cite{13}; (v) study of the nonlinear dynamical
evolution of a driven two-photon Jaynes-Cummings model \cite{14}; (vi) the
study of a generalized Jaynes-Cummings models, including dissipation \cite%
{15,16,16a} and multiphoton interactions \cite{17,bb1}; etc. In all these
cases, with interest either on the field or on atomic properties, the
theoretical approach traditionally assumes the atom-field coupling as a
constant parameter. Comparatively, the number of works in the literature is
very small when one considers such coupling and the atomic frequency as time
dependent parameters \cite{law,l1,l22,l2,l3,13.1}, including time dependent
amplitudes \cite{abdalla}. However, this scenario is also relevant; \ for
example, the state of two qubits (qubits stand for quantum bits) with a
desired degree of entanglement can be generated via a time dependent
atom-field coupling \cite{olaya}. Actually, such coupling can modify the
dynamical properties of the atom and the field, with transitions that
involve a large number of photons \cite{yang}. In general, these studies are
simplified by neglecting the atomic decay from an excited level. Theoretical
treatments taking into account this complication of the real world may
employ a modified JCM. In these case, as expected, the state describing the
system decoheres, since the presence of dissipation destroys the state of a
system as time flows.

In the present work we extend what we have learned from the JCM applied to
the atom-field interaction to investigate a more advantageous system, from
the experimental viewpoint (faster response, better controllability, and
useful scalability for quantum computation \cite{y1} ) by considering\ a
nanomechanical resonator (NR) interacting with a Cooper pair box (CPB) \cite%
{17,16,18}. Such nanodevice has been explored extensively in the literature,
e.g., to investigate: (i) quantum nondemolition measurements \cite{ak1,e1},
(ii) decoherence of nonclassical states, as Fock states and superposition
states describing mesoscopic systems \cite{e2,c2}, etc. The fast advance in
the technique of fabrication in nanotechnology has implied great interest in
the study of the NR system in view of its potential applications, as a
sensor - to be used in biology, astronomy, quantum computation, and in
quantum information \cite{ak2,bb2,bb3}, to implement the quantum qubit \cite%
{ak3} and in the production of nonclassical states, as Fock state \cite%
{akk,bb4}, Schr\"{o}dinger's cat state \cite{akk2}, squeezed states \cite%
{a34}, clusters states \cite{ak}, etc. In particular, when accompanied by
superconducting charge qubits, the NR has been used to prepare entangled
states \cite{akk1}. Zhou et al.\cite{a34} have proposed a scheme to prepare
squeezed states using a NR coupled to a CPB qubit; in this proposal the
NR-CPB coupling is under an external control while the connection between
these two subsystems plays an important role in quantum computation. Such a
control is achieved via convenient change of the system parameters, which
can set ``on" and ``off" the interaction between the NR and the CPB, on
demand.

In this report we will investigate the CPB excitation inversion, its
control, and the average photon number in the NR. We will consider
dissipation in the CPB due to a decay rate from excited to ground states. We
will also verify in which way the time dependence of the CPB-NR coupling
modifies these two properties. To this end we must solve the time evolution
of the whole CPB-NR system, via the approach presented in the following
Section.

\section{Model Hamiltonian for the CPB-NR system}

\begin{figure}[tbh]
\centering  
\includegraphics[width=10cm, height=12cm]{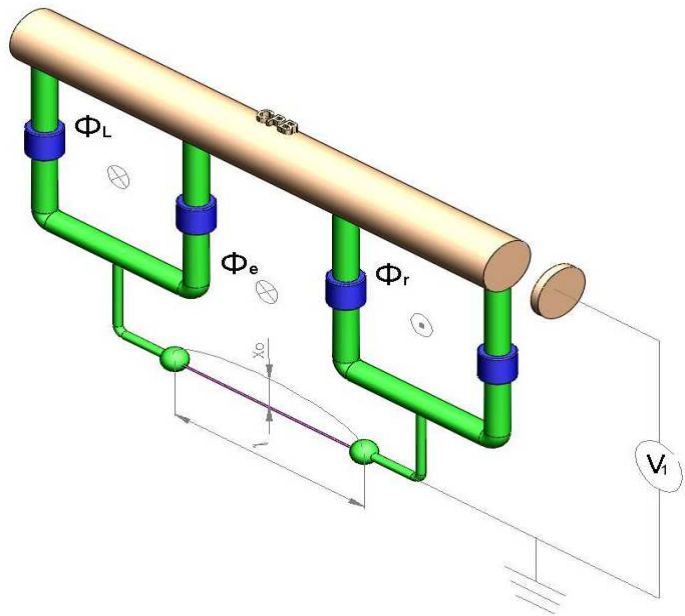}
\caption{\textit{Model for the CPB-NMR coupling.}}
\label{cooper}
\end{figure}

A Josephson charge qubit system has been used to couple with a NR. Here we
study a modified model where a CPB is coupled to a NR, as shown in Fig. \ref%
{cooper} below. The scheme is inspired by the works of Jie-Qiao Liao et al.
\cite{ak3} and Zhou et al. \cite{a34} where we have substituted each
Josephson junction by two of them. This creates a new configuration that
includes a third loop. A superconducting CPB charge qubit is adjusted via a
voltage $V_{1}$\ at the system input and a capacitance $C_{g}$. We want that
the scheme attains an efficient tunneling effect for the Josephson energy.
In Fig. \ref{cooper} we observe three loops: one great loop between two
small ones. This makes it easier controlling the external parameters of the
system since the control mechanism includes the input voltage $V_{1}$ plus
three external fluxes $\Phi _{L},$ $\Phi _{r}$ and $\Phi _{e}(t)$. In this
way one can induce small neighboring loops. The great loop contains a NR
which is modeled as a harmonic oscillator with a high-Q mode of frequency $%
\Omega $ and its effective area in the center of the apparatus changes as
the NR oscillates, which creates an external flux $\Phi _{e}(t)$ that
provides the CPB-NR coupling.

In pursuing the quantum behavior of a macro scale object the nano scale
mechanical resonator plays an important role. At sufficiently low
temperature the zero-point fluctuation of the NR will be comparable to its
thermal Brownian motion. The detection of zero-point fluctuations of the NR
can give a direct test of Heisenberg's uncertainty principle. With a
sensitivity up to 10 times the amplitude of the zero-point fluctuation,
LaHaye et al \cite{lahaye} have experimentally detected the vibrations of a
20 MHz mechanical beam of tens of micrometres size. For a 20MHz mechanical
resonator its temperature must be cooled below 1mK to suppress the thermal
fluctuation. For a GHz mechanical resonator a temperature of 50mK is
sufficient to effectively freeze out its thermal fluctuation and let it
enter the quantum regime. This temperature is already attainable in dilution
refrigerators.

%
%
In this work we will assume the four Josephson junctions being identical,
with the same Josephson energy $E_{J}^{0}$, the same being assumed for the
external fluxes $\Phi _{L}$ and $\Phi _{r}$, i.e., with same magnitude but
opposite sign: $\Phi _{L}=-\Phi _{r}=\Phi _{x}$. This interaction actually
couples the two subsystems. Together with the free Hamiltonian of flux qubit
and NR, the Hamiltonian of the whole system reads

\begin{equation}
\hat{H}=\Omega \hat{a}^{\dagger }\hat{a}+4E_{c}\left( N_{g}-\frac{1}{2}%
\right) \hat{\sigma}_{z}-4E_{J}^{0}\cos \left( \frac{\pi \Phi _{x}}{\Phi _{0}%
}\right) \cos \left( \frac{\pi \Phi _{e}}{\Phi _{0}}\right) \hat{\sigma}_{x},
\label{a1}
\end{equation}%
where $\hat{a}^{\dagger }(\hat{a})$ is the creation (annihilation) operator
for the NR excitation, with frequency $\Omega $ \ and mass $m$; $E_{J}^{0}$
and $E_{c}$ are respectively the energy of each Josephson junction and the
charge energy of a single electron; $C_{g}$ and $C_{J}^{0}$ are the input
capacitance and the capacitance of each Josephson tunnel, respectively. $%
\Phi _{0}=h/2e$ is the quantum flux and $N_{g}=C_{g}V_{1}/2e$ is the charge
number in the input with the input voltage $V_{1}$. We have used the Pauli
matrices to describe our system operators, where the states $\left\vert
0\right\rangle $ and $\left\vert 1\right\rangle $ represent the number of
extra Cooper pairs in the superconducting island. We have: $\hat{\sigma}%
_{z}=\left\vert 0\right\rangle \left\langle 0\right\vert -\left\vert
1\right\rangle \left\langle 1\right\vert $, $\hat{\sigma}_{x}=\left\vert
0\right\rangle \left\langle 1\right\vert -\left\vert 1\right\rangle
\left\langle 0\right\vert $ and $E_{C}=e^{2}/\left( C_{g}+4C_{J}^{0}\right)
. $

The magnetic flux can be written as the sum of two terms,
\begin{equation}
\Phi _{e}=\Phi _{1}+B\ell \hat{x},  \label{a4}
\end{equation}%
where the first term $\Phi _{1}$ is the induced flux, corresponding to the
equilibrium position of the NR and the second term describes the
contribution due to the vibration of the NR; $B$ represents the magnetic
field created in the loop. We have assumed the displacement $\hat{x}$
described as $\hat{x}=x_{0}(\hat{a}^{\dagger }+\hat{a})$, where $x_{0}=\sqrt{%
m\Omega /2}$ is the amplitude of the oscillation. Substituting the Eq.(\ref%
{a4}) in Eq.(\ref{a1}) and controlling the flux $\Phi _{1}$ we can adjust $%
\cos \left( \frac{\pi \Phi _{1}}{\Phi _{0}}\right) =0$, and making the
approximation $\pi B\ell x/\Phi _{0}<<1$ the above Hamiltonian results as
(in rotating wave approximation),
\begin{equation}
\hat{H}=\Omega \hat{a}^{\dagger }\hat{a}+\frac{1}{2}\omega _{0}\hat{\sigma}%
_{z}+\lambda _{0}(\hat{\sigma}_{+}\hat{a}+\hat{a}^{\dagger }\hat{\sigma}%
_{-}),
\end{equation}%
where the constant coupling $\lambda _{0}=-4E_{J}^{0}\cos \left( \frac{\pi
\Phi _{x}}{\Phi _{0}}\right) \left( \frac{\pi B\ell x_{0}}{\Phi _{0}}\right)
$ and the effective energy $\omega _{0}=8E_{c}\left( N_{g}-\frac{1}{2}%
\right) .$ An important advantage of this coupling mechanism is its easy and
convenient controllability.

Next, we will extend the previous approach to a more general scenario by
substituting $\Omega \rightarrow \omega (t)=\Omega +f\left( t\right) $ and $%
\lambda _{0}\rightarrow \lambda (t)=\lambda _{0}\left[ 1+f\left( t\right)
/\Omega \right] $ \cite{yang,c1,jf}; in addition we assume the presence of a
constant decay rate $\gamma $ in the CPB; $\omega _{0}$ is the transition
frequency of the CPB and $\lambda _{0}$ stands for the CPB-NR\ coupling. $%
\hat{\sigma}_{\pm }$ and $\hat{\sigma}_{z}$ are the CPB transition and
excitation inversion operators, respectively; they act on the Hilbert space
of atomic states and satisfy the commutation relations $\left[ \hat{\sigma}%
_{+},\hat{\sigma}_{-}\right] =\hat{\sigma}_{z}$ and $\left[ \hat{\sigma}_{z},%
\hat{\sigma}_{\pm }\right] =\pm \hat{\sigma}_{\pm }$. As well known, the
coupling parameter $\lambda (t)$ is proportional to $\sqrt{\omega
(t)/V\left( t\right) }$, where the time dependent quantization volume $%
V\left( t\right) $ takes the form $V\left( t\right) =V_{0}/\left[ 1+f\left(
t\right) /\Omega \right] $ \cite{scully,l2,jf}. Accordingly, we obtain the
new (\textit{non hermitian}) Hamiltonian

\begin{equation}
\hat{H}=\omega (t)\hat{a}^{\dagger }\hat{a}+\frac{1}{2}\omega _{0}\hat{\sigma%
}_{z}+\lambda (t)(\hat{\sigma}_{+}\hat{a}+\hat{a}^{\dagger }\hat{\sigma}%
_{-})-i\frac{\gamma }{2}\left\vert 1\right\rangle \left\langle 1\right\vert .
\label{b1}
\end{equation}

It is worth remembering that non Hermitian Hamiltonians (NHH) are largely
used in the literature. To give some few examples we mention: Ref. \cite{nh5}%
, where the authors use a NHH and an algorithm to generalize the
conventional theory; Ref. \cite{nh1}, using a NHH to get information about
entrance and exit channels; Ref. \cite{nh6}, using non Hermitian techniques
to study canonical transformations in quantum mechanics; Ref. \cite{nh7},
solving quantum master equations in terms of NHH; Ref. \cite{nh3}, using a
new approach for NHH to study the spectral density of weak H-bonds involving
damping; Ref. \cite{nh8}, studying NHH with real eigenvalues; Ref. \cite{nh4}%
, using a canonical formulation to study dissipative mechanics exhibing
complex eigenvalues;\ Ref. \cite{nh9}, studying NHH in non commutative
space, and more recently: Ref. \cite{nh10}, studying the optical realization
of relativistic NHH; Ref. \cite{l2}, studying the evolution of entropy of
atom-field interaction; Ref. \cite{l22}, using a damping JC-Model to study
entanglement between two atoms, each of them lying inside different cavities

\section{Solving the CPB-NR system}

The state that describes this time dependent system can be written in the
form

\begin{equation}
\left\vert \Psi \left( t\right) \right\rangle =\sum\nolimits_{n=0}^{\infty
}(C_{0,n}\left( t\right) \left\vert 0,n\right\rangle +C_{1,n}\left( t\right)
\left\vert 1,n\right\rangle ),  \label{b2}
\end{equation}%
where $\left\vert 0,n\right\rangle $ $\left( \left\vert 1,n\right\rangle
\right) $ represents the CPB in its excited state $\left\vert 1\right\rangle
$ (ground state $\left\vert 0\right\rangle $). Taking the CPB initially
prepared in its excited state $\left\vert 1\right\rangle $ and the NR in a
coherent states $\left\vert \alpha \right\rangle $, and expanding coherent
state component in the Fock's basis, i.e., $\left\vert \alpha \right\rangle
=exp(-|\alpha |^{2}/2)\sum_{n=o}^{\infty }(\alpha ^{n}/\sqrt{n!})|n\rangle $%
,\ we have $\left\vert \alpha \right\rangle =\sum\nolimits_{n=0}^{\infty
}F_{n}\left\vert n\right\rangle $.\ Assuming the NR and CPB decoupled at $%
t=0 $ and the initial conditions $C_{0,n}\left( 0\right) =0$ and $%
\sum\nolimits_{n=0}^{\infty }\left\vert C_{1,n}\left( 0\right) \right\vert
^{2}=1$ we may write the Eq. (\ref{b2}) as $\left\vert \Psi \left( 0\right)
\right\rangle =\sum\nolimits_{n=0}^{\infty }F_{n}\left\vert 1,n\right\rangle
.$

The time dependent Schr\"{o}dinger equation for this system is

\begin{equation}
i\frac{d\left\vert \Psi \left( t\right) \right\rangle }{dt}=\hat{H}%
\left\vert \Psi \left( t\right) \right\rangle ,  \label{b5}
\end{equation}%
with the Hamiltonian $\hat{H}$ given in Eq. (\ref{b1}).\emph{\ }Substituting
Eq.(\ref{b1}) in Eq.(\ref{b5}) we get the (coupled) equations of motion for
the probabilitity amplitudes $C_{1,n}(t)$ and $C_{0,n+1}(t)$:%
\begin{eqnarray}
\frac{\partial C_{1,n}(t)}{\partial t} &=& -in\omega (t)C_{1,n}(t)-\frac{i}{2%
}\omega _{0}C_{1,n}(t)-i\lambda (t)\sqrt{n+1}C_{0,n+1}(t)-\frac{\gamma }{2}%
C_{1,n}(t),\phantom {aaaaaa}  \label{b8} \\
\frac{\partial C_{0,n+1}(t)}{\partial t}&=&-i(n+1)\omega (t)C_{0,n+1}(t)+%
\frac{i}{2}\omega _{0}C_{0,n+1}(t)-i\lambda (t)\sqrt{n+1}C_{1,n}(t).
\label{b9}
\end{eqnarray}%
The numerical solutions of the coefficients $C_{1,n}(t)$, $C_{0,n+1}(t)$
furnish the quantum dynamical properties of the system, including the CPB-NR
entanglement.

As well known, in the presence of decay rate $\gamma $ in the CPB the state
of the whole CPB-NR system becomes mixed. In this case its description
requires the use of the density operator $\hat{\rho}_{CN}$, which describes
the entire system. To obtain the reduced \ density matrix describing the CPB
(NR) subsystem we must trace over variables of the NR (CPB) subsystem. For
example, $\hat{\rho}_{NR}=Tr_{CPB}(\hat{\rho}_{CN})$:
\begin{equation}
\hat{\rho}_{NR}=\sum\nolimits_{n,n^{\prime }=o}^{\infty }\left[
C_{1,n}(t)C_{1,n^{\prime }}^{\ast }(t)+C_{0,n}(t)C_{0,n^{\prime }}^{\ast }(t)%
\right] \left\vert n\right\rangle \left\langle n\prime \right\vert .
\label{c2}
\end{equation}

\section{Excitation Inversion of the CPB}

The CPB excitation inversion, here denoted as $I(t)_{CPB}$, is an important
observable of two level systems. It is defined as the difference of
probabilities of finding the system in the excited and ground state, as
follows%
\begin{equation}
I(t)_{{CPB}}=\sum\nolimits_{n=0}^{\infty }\left[ \left\vert
C_{1,n}(t)\right\vert ^{2}-\left\vert C_{0,n+1}(t)\right\vert ^{2}\right] .
\label{joao}
\end{equation}

\begin{figure}[tbh]
\centering  
\includegraphics[width=9cm, height=8cm]{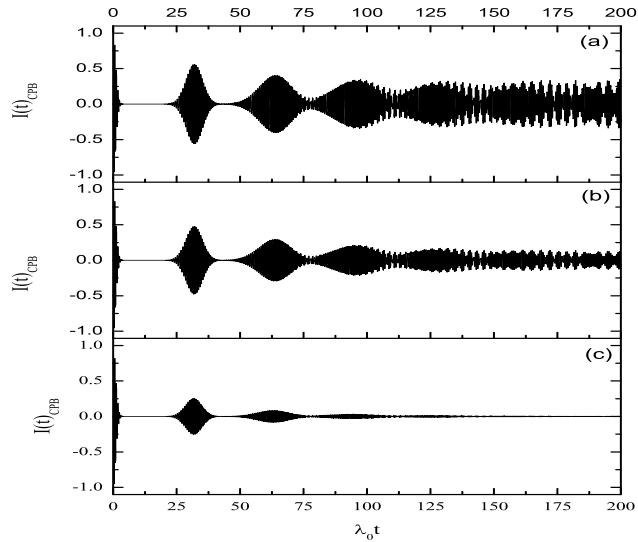}
\caption{\textit{Time evolution of the Excitation Inversion in the CPB with $%
\left\langle n\right\rangle =25$, $\Omega =\protect\omega _{0}=2000\protect%
\lambda _{0}$, for $f(t)=0$ (resonance) and different values of decay rates $%
\protect\gamma$: (a) $\protect\gamma =0.01\protect\lambda _{0}$, (b) $%
\protect\gamma =0.05\protect\lambda _{0}$ and (c) $\protect\gamma =0.5%
\protect\lambda _{0}$.}}
\label{inversao1}
\end{figure}

\begin{figure}[tbh]
\centering  
\includegraphics[width=9cm, height=8cm]{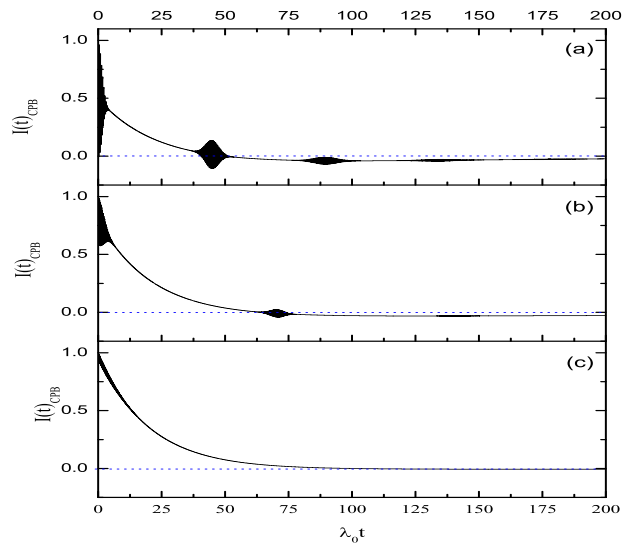}
\caption{\textit{Same as in Fig. \protect\ref{inversao1} for $\protect\gamma %
=0.05\protect\lambda _{0}$ and different values of detunings $($cf. $%
f(t)=\Delta =const.)$: (a) $\Delta =10\protect\lambda _{0}$\ , (b) $\Delta
=20\protect\lambda _{0}$, (c) $\Delta =60\protect\lambda _{0}$.}}
\label{inversao2}
\end{figure}

\begin{figure}[tbh]
\centering  
\includegraphics[width=9cm, height=8cm]{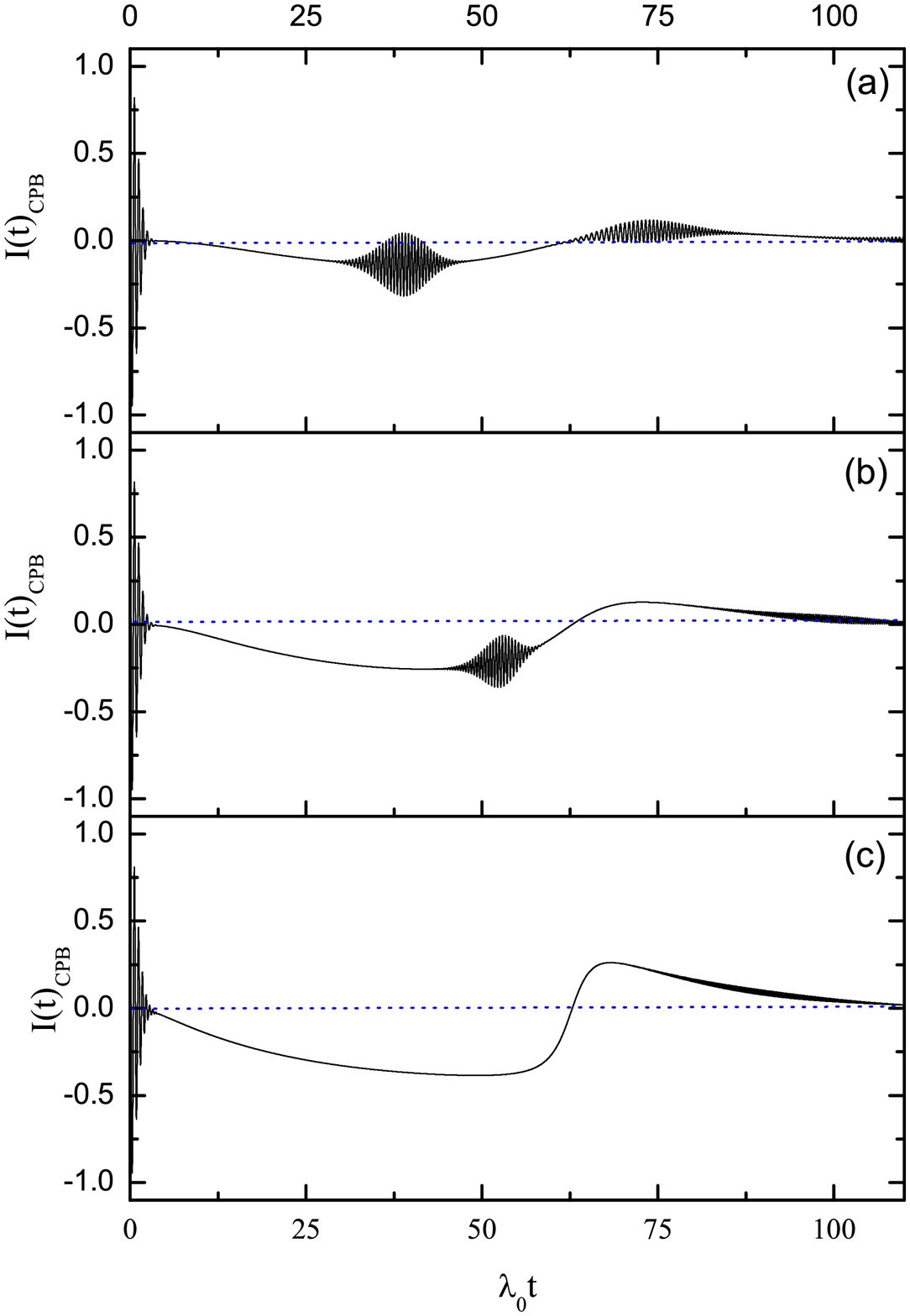}
\caption{Same as in Fig. \protect\ref{inversao1} for $\protect\gamma =0.05%
\protect\lambda _{0}$ and different time-dependent detunings\textit{\ (cf. $%
f(t)=\protect\eta \sin (\protect\omega \prime t)$): (a) $\protect\eta=10%
\protect\lambda _{0}$ and $\protect\omega \prime =0.05\protect\lambda _{0}$,
(b) $\protect\eta=20\protect\lambda _{0}$ and $\protect\omega \prime =0.05%
\protect\lambda _{0},$ (c) $\protect\eta=60\protect\lambda _{0}$ and $%
\protect\omega \prime =0.05\protect\lambda _{0}.$}}
\label{inversao3}
\end{figure}

The Eq. (\ref{joao}) allows one to look at the time evolution of the CPB
excitation inversion. First, we assume the resonant case $(f(t)=0)$ for
different values of the decay rate $\gamma $, with $\alpha =5$ and $\Omega
=\omega _{0}=2000\lambda _{0}$ and assuming the NR initially in a coherent
state with the average number of excitations $\left\langle n\right\rangle
=25 $ as shown in Fig. \ref{inversao1}. With the exceptions of amplitudes,
the plots (a), (b) and (c) of Fig. \ref{inversao1} show identical for
collapse-revival effects: the higher the decay rate the lower is the
amplitude of oscillation. In presence of detuning, where $f(t)=\Delta =const$%
, where $\Delta \ll \omega _{0}$, $\Omega $, we see that the excitation
inversion in Fig. \ref{inversao2}(a) occurs within the interval $30<\lambda
_{0}t<50$, whereas in Fig. \ref{inversao2}(b) it occurs in the range $%
60<\lambda _{0}t<75$ and in Fig. \ref{inversao2}(c) the excitation inversion
goes to zero rapidly. Considering the case of variable detuning, $f(t)=\eta
\sin (\omega \prime t)$, we see in the plots (a), (b) and (c) of Fig. \ref%
{inversao3} that the excitation inversion occurs frequently in Fig. \ref%
{inversao3}(a) and desapears when the parameter $\eta $ increases, as shown
in the plots (b) and (c). Now, even considering the worst results obtained
in the off-resonant cases, with detuning $\Delta =\eta =60\lambda _{0}$, as
shown in Fig. \ref{inversao2}(c) and Fig. \ref{inversao3}(c) we see that we
can recover the collapse-revival effects via the increase of the parameter $%
\omega \prime ,$ as shown in the plots (a), (b) and (c) of Fig. \ref%
{inversao4}. So, the parameter $\omega \prime $ plays a fundamental role in
the control of collapse and revival effect.

\begin{figure}[tbh]
\centering  
\includegraphics[width=9cm, height=8cm]{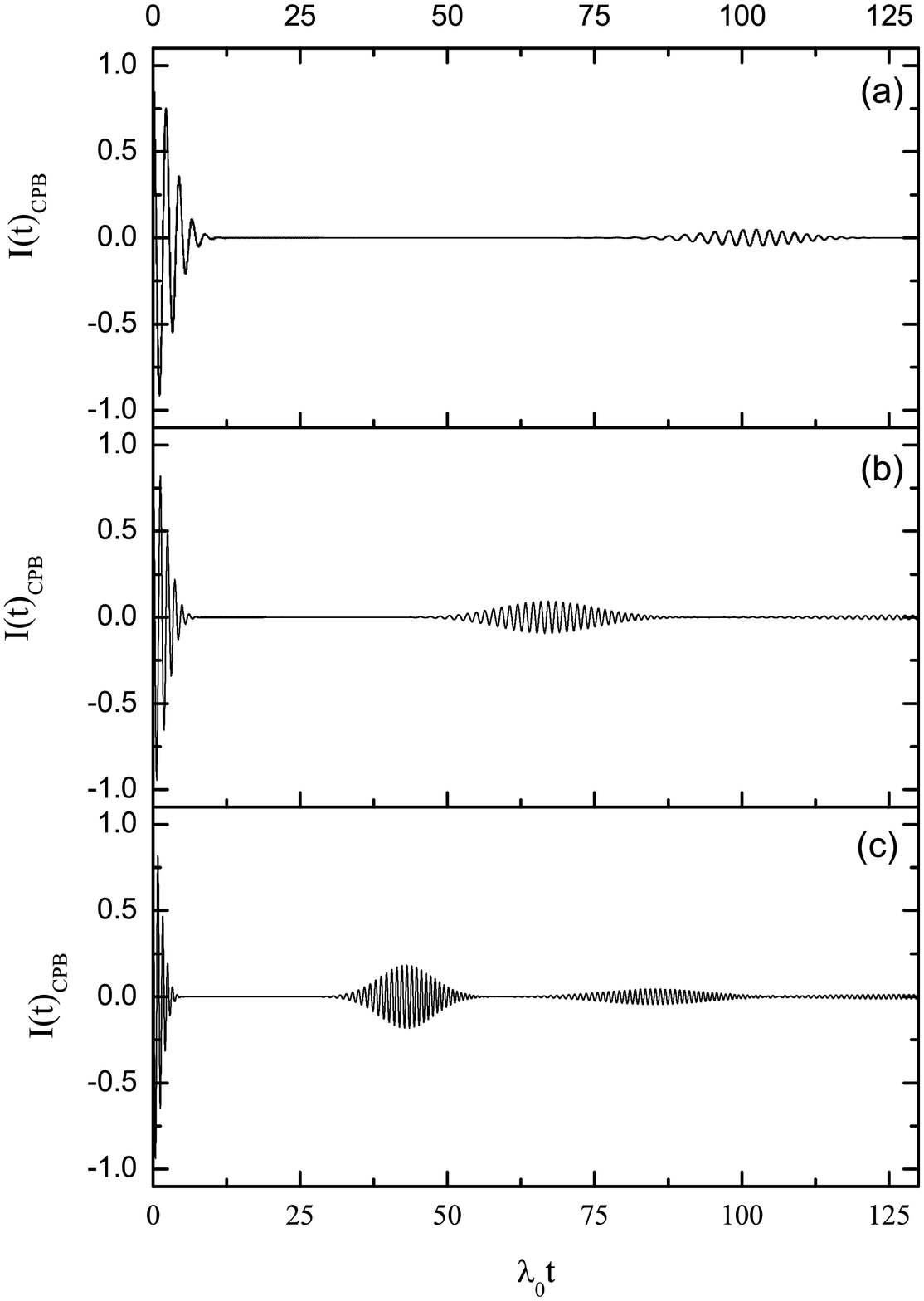}
\caption{Same as in Fig. \protect\ref{inversao1} for $\protect\gamma =0.05%
\protect\lambda _{0}$ and different time-dependent detunings\textit{\ (cf. $%
f(t)=\protect\eta\sin (\protect\omega \prime t)$): (a) $\protect\eta=60%
\protect\lambda _{0}$ and $\protect\omega \prime =20\protect\lambda _{0}$,
(b) $\protect\eta=60\protect\lambda _{0}$ and $\protect\omega \prime =40%
\protect\lambda _{0},$ (c) $\protect\eta=60\protect\lambda _{0}$ and $%
\protect\omega \prime =58\protect\lambda _{0}$.}}
\label{inversao4}
\end{figure}

\begin{figure}[tbh]
\centering  
\includegraphics[width=9cm, height=8cm]{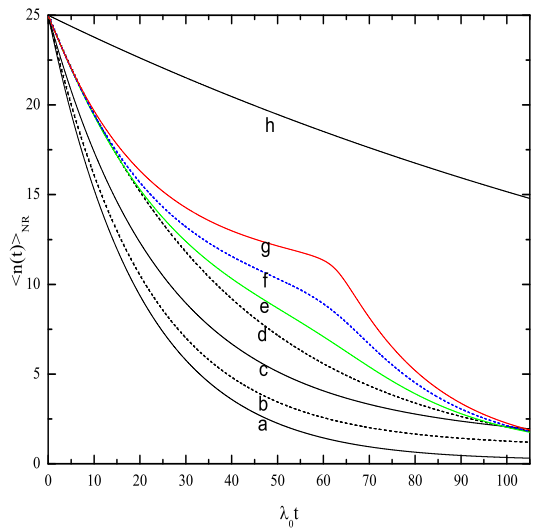}
\caption{\textit{Average number of excitations of the NR, versus time, for $%
\left\langle n(0)\right\rangle =25$, $\Omega =\protect\omega _{0}=2000%
\protect\lambda _{0}$, for the following cases: (a) off-resonance }$\protect%
\gamma =0.05\protect\lambda _{0}$, $\Delta =60\protect\lambda _{0},$\textit{%
\ (b) off-resonance }$\protect\gamma =0.05\protect\lambda _{0}$, $\Delta =20%
\protect\lambda _{0},$\textit{\ (c) off-resonance }$\protect\gamma =0.05%
\protect\lambda _{0}$, $\Delta =10\protect\lambda _{0},$\textit{\ \ (d) on
resonance }$\protect\gamma =0.5\protect\lambda _{0},$\textit{\ (e)
off-resonance }$\protect\gamma =0.05\protect\lambda _{0},$ $\protect\eta=10%
\protect\lambda _{0}$ and $\protect\omega \prime =0.05\protect\lambda _{0},$%
\textit{\ (f) off-resonance }$\protect\gamma =0.05\protect\lambda _{0},$ $%
\protect\eta=20\protect\lambda _{0}$ and $\protect\omega \prime =0.05\protect%
\lambda _{0},$\textit{(g) off-resonance }$\protect\gamma =0.05\protect%
\lambda _{0},$ $\protect\eta=60\protect\lambda _{0}$ and $\protect\omega %
\prime =0.05\protect\lambda _{0},$\textit{\ (h) on resonance }$\protect%
\gamma =0.05\protect\lambda _{0}$. }
\label{nmf1}
\end{figure}

Fig. \ref{nmf1} shows plots of the mean value of the NR excitations in the
presence of CPB decay rate for various values of amplitude of oscillations
(parameter $\eta $). Plots (\textbf{d)} and (\textbf{h)} are for the
resonant case: in (\textbf{d) }the decay rate is greater than in (\textbf{h)}%
. Plots (\textbf{a)}, (\textbf{b)}, and (\textbf{c)} are for constant decay
rates, with (time independent) detuning that increases from (\textbf{c) }$%
\rightarrow $\textbf{\ (b) }$\rightarrow $\textbf{\ (a)}. Finally, plots (%
\textbf{e)}, (\textbf{f)} and (\textbf{g)} are for time dependent detuning,
with the parameter $\eta $ increasing from (\textbf{e) }$\rightarrow $%
\textbf{\ (f) }$\rightarrow $\textbf{\ (g)}. \ We note that the three plots
for time dependent detunings (\textbf{e)}, (\textbf{f)} and\textit{\ (}%
\textbf{g}) are better than those for constant detunings (\textbf{a)}, (%
\textbf{b)} and (\textbf{c}): despite all plots are concerned with the same
decay rate, the first group is more robust against decay. For example,
comparing the plots (\textbf{a}) and (\textbf{g)}: although in \textbf{(a)}
the fixed detuning is $\Delta =60\lambda _{0}$ and in (\textbf{g)} maximum
detuning is $\Delta _{\max }=\eta =60\lambda _{0}$ we see that in the last
case the average value of the NR decays more slowly. One observes in the
plot (\textbf{c)} of Fig. \ref{inversao4} that the interval $\lambda _{0}t$,
where the presence of the time dependent detuning recovers the
collapse-revival effect, coincides with that in the plot (\textbf{g)} of
Fig. \ref{nmf1} where the mean value of excitation is around 5 times greater
than in the case of constant detuning (plot (\textbf{a)} of same figure).


\section{Conclusion}

We have considered a Hamiltonian model that describes a CPB-NR interacting
system to study the CPB\ excitation inversion, $I(t)_{{CPB}},$ and the
average excitation number of the NR, $\left\langle n(t)\right\rangle _{NR}$.
We have also considered the off-resonant case, with various values of the
detuning parameter $f\ $($f=0;f=\Delta ;$ and $f=\eta \sin (\omega \prime t)$%
) and in the presence of CPB\ decay (about 10 times greater than the
(neglected) NR decay). These properties are characteristics of the entangled
state that describes this coupled system for various values of the
parameters involved. We have assumed the CPB\ initially in its excited state
and the NR initially in a coherent state (see preparation in \cite{17}). So,
the following three scenarios were treated: (\textit{i}) both subsystems in
resonance (detuning $f=0$); (\textit{ii}) off-resonance, with a constant
detuning $(f=\Delta \neq 0)$, and (\textit{iii}) with a time dependent
detuning $(f(t)=\eta \sin (\omega \prime t))$. The results were discussed in
the previous Section: in resume, concerning the CPB excitation inversion, an
interesting result emerged: although the presence of a constant detuning
destroys the collapse and revivals of the excitation inversion, these
effects are restituted by the action of convenient time dependent detunings
- even in the presence of damping in the CPB; concerning the NR average
excitation number, another interesting result appeared: convenient choices
of the time dependent detuning $f(t)$ makes the NR subsystem more robust
against the decay affecting the CPB subsystem. For constant values of
detuning, our numerical results are similar to others in the literature
using a master equation (see, e.g., Ref. \cite{16}).

Finally we emphasize that the change in magnetic flux $\Phi_{e}$ (cf. Fig. %
\ref{cooper}), due to the presence of an external force upon the NR, is the
responsible for controlling the parameters $\omega (t)$ and $\lambda (t)$.

\section{Acknowledgements}

The authors thank the FAPEG (CV) and CNPq (ATA, BB), Brasilian Agencies, for
partially supporting this paper.


\end{document}